\documentstyle[]{article}
\font\cero=cmss10 scaled 1728
\font\uno=cmssbx10 scaled 1200
\begin{document}
\small{
\begin{flushleft}
{\cero Towards a covariant canonical formulation for closed
topological defects without boundaries} \\[3em]
\end{flushleft}
{\sf R. Cartas-Fuentevilla}\\
{\it Instituto de F\'{\i}sica, Universidad Aut\'onoma de Puebla,
Apartado postal J-48 72570, Puebla Pue., M\'exico.}
(rcartas@sirio.ifuap.buap.mx) \\[4em]

 On the basis of the covariant description of the canonical
formalism for quantization, we present the basic elements of the
symplectic geometry for a restricted class of topological defects
propagating on a curved background spacetime. We discuss the
future extensions of the present results. \\

\noindent Running title:  Towards a covariant....\\

\begin{center}
{\uno I. INTRODUCTION}
\end{center}
\vspace{1em}

It is undeniable the current importance of the study of extended
objects in physics, mainly in many high energy physics contexts.
Such solitonic structures emerge, for example, in the context of
cosmological phase transitions in the early universe, and playing
a fundamental role at the Planck scale, where
(super)string/M-theory drives a revolution without precedent in
physics, such as to provide a quantum formulation for gravity.
However, a great variety of aspects and questions are still not
complete and properly understood, both at classical and quantum
level, included the transition between the classical and quantum
domains. This happens even for the simplest prototype for a
extended object, the Dirac-Nambu-Goto model.

Within the general context of quantization, the canonical
formalism is based in a Hamiltonian scheme that traditionally
requires an explicit singling out of the time as evolution
variable, which implies the space-time to be topologically a
direct product of space and time, as opposed to the general
covariance imposed by the relativity theory. However, these
difficulties have been overcame by means of a generalized
Hamiltonian scheme in which space and time dimensions are treated
on equal footing, leading up to a canonical formalism manifestly
covariant ([1] and references therein). This covariant formalism
has been applied for the analysis of field theories such as
Yang-Mills and General Relativity, originally in Ref.\ [1] by
Witten {\it et al}, and more recently, incorporating the adjoint
operators formalism, by ourselves in Ref.\ [2], open superstrings
\cite{3}, the Wess-Zumino-Witten model \cite{4}, two-dimensional
gravity models \cite{5}, among other ones; however, it has not
been applied for the analysis of topological defects of arbitrary
dimensions. That is due to several reasons: first, the procedures
previously employed for field theories can not be extended
directly for the natural geometrical quantities that describe the
dynamics of a defect, and the translation turns out to be awkward
and obscure. Additionally, within that covariant canonical
formalism, one requires precisely of a {\it covariant}
description of the geometry of deformations of the subject under
study, which, in the case of topological defects (unlike field
theories), is hardly in current development, and is yet not well
understood. Attempts for overcoming such limitations include, for
example, the formalism developed by Capovilla and Guven (CG) for
the case of closed topological defects without boundaries
\cite{6}, and subsequently for extended objects with edges
\cite{7}. Hence, the purpose of the present study is to employ the
CG formalism  for developing those elements which allow to
establish a covariant quantization scheme described above, for
topological defects, particularly those governed by a higher
dimensional generalization of the Dirac-Nambu-Goto action (or
branes), on a curved spacetime. We shall focus our attention, in
this preliminary effort, on closed defects without physical
boundaries.

In the covariant canonical formalism that we attempt to apply, the
knowledge of the so called covariant space phase is crucial.
Therefore, in Section IV , we outline the definition of such
space, and the exterior calculus associated with. Although such
definitions and concepts come entirely from Ref.\ \cite{1}, they
will be suitably adjusted to the treatment of topological defects.
In Section V, we incorporate the (self)-adjoint operators
formalism, which allows us to obtain a {\it covariantly}
conserved bilinear form on the phase space, which in turns, will
be associated directly with the wanted symplectic form. In
Section VI, a symplectic structure with certain invariance
properties is explicitly constructed on the phase space. We have
no conclusions in this letter, instead we have a lot of open
questions for future works, and they are discussed in Section
VII.  In next section, we summarize the geometry of the embedding
of the brane worldsheet in a background spacetime and of their
deformations given in Reference \cite{6}. \\[2em]

\noindent {\uno II. GEOMETRY OF THE EMBEDDING AND THEIR
DEFORMATIONS}
\vspace{1em}

\noindent{\uno 2.1 The embedding}
\vspace{.5cm}

The $D$-dimensional brane dynamics is usually given by a oriented
timelike worldsheet {\it m} described by the embedding functions
$x^{\mu} = X^{\mu} (\xi^{a})$, $\mu = 0, ..., N - 1$ and $ a = 0,
..., D$, in a $N$-dimensional ambient spacetime {\it M} endowed
with the metric $g_{\mu\nu}$. Such functions specify the
coordinates of the brane, and the $\xi^{a}$ correspond to internal
coordinates on the worldsheet.

At each point of {\it m}, $e_{a} \equiv X^{\mu}_{,a}
\partial_{\mu} \equiv e^{\mu}_{a} \partial_{\mu}$, generate a basis
of tangent vectors to {\it m}; thus, the induced
$(D+1)$-dimensional worldsheet metric is given by $\gamma_{ab} =
e^{\mu}_{a} e^{\nu}_{b} g_{\mu\nu} = g(e_{a}, e_{b})$.
Furthermore, the $(N - D)$ vector fields $n^{i}$ normal to {\it
m}, are defined by
\begin{equation}
     g(n^{i}, n^{j}) = \delta^{ij}, \quad g(e_{a}, n^{i}) = 0.
\end{equation}
Tangential indices are raised and lowered by $\gamma^{ab}$ and
$\gamma_{ab}$, respectively, whereas normal vielbein indices by
$\delta^{ij}$ and $\delta_{ij}$ respectively, and this fact will
be used implicitly below. The collection of vectors $\{e_{a},
n^{i}\}$, which can be used as a basis for the spacetime vectors,
satisfies the generalized Gauss-Weingarten equation:
\[
   D_{a} e_{b} = \gamma_{ab} {^{c}}e_{c} - K_{ab} {^{i}}n_{i},
\quad D_{a} n_{i} = K_{ab} {^{i}}e^{b} + \omega_{a} {^{ij}}n_{j},
\]
where $D_{a} \equiv e^{\mu}_{a} D_{\mu}$ ($D_{\mu}$ is the
torsionless covariant derivative associated with $g_{\mu\nu}$);
thus, the connection coefficients $\gamma_{ab} {^{a}}$ compatible
with $\gamma_{ab}$ is given by $\gamma_{ab} {^{c}} = g(D_{a}
e_{b}, e^{c}) = \gamma_{ba} {^{c}}$, and the {\it i}th extrinsic
curvature of the worldsheet  by $K_{ab} {^{i}} = - g (D_{a} e_{b},
n^{i}) = K_{ba}{^{i}}$. Similarly the extrinsic twist potential of
the worldsheet is defined by $\omega_{a} {^{ij}} = g(D_{a} n^{i},
n^{j}) = - \omega_{a} {^{ji}}$. Such a potential allows us to
introduce a worldsheet covariant derivative
$(\widetilde{\nabla}_{a})$ defined on fields $(\Phi^{i} {_{j}})$
transforming as tensors under normal frame rotations:
\begin{equation}
     \widetilde{\nabla}_{a} \Phi^{i} {_{j}} \equiv \nabla_{a}
\Phi^{i} {_{j}} - \omega_{a} {^{ik}} \Phi_{k j} - \omega_{ajk}
\Phi^{ik},
\end{equation}
where $\nabla_{a}$ is the (torsionless) covariant derivative
associated with $\gamma_{ab}$. \\[2em]

\noindent{\uno 2.2 Deformations of the intrinsic geometry}
\vspace{.5cm}

The physically observable measure of the deformation of the
embedding {\it m} is given by the orthogonal projection of the
infinitesimal spacetime variation $\xi^{\mu} \equiv \delta X^{\mu}
= n^{\mu}_{i} \phi^{i}$, characterized by $N-D$ scalar fields
$\phi^{i}$. Defining the vector field $\delta \equiv n_{i}
\phi^{i}$, the displacement induced in the tangent basis $\{ e_{a}
\}$ along $\delta$ depends on $\phi^{i}$ and on their first
derivatives:
\begin{equation}
     D_{\delta} e_{a} = \beta_{ab} e^{b} + J_{ai}n^{i},
\end{equation}
where $D_{\delta} \equiv \delta^{\mu} D_{\mu}$, and
\begin{equation}
     \beta_{ab} = g(D_{\delta} e_{a}, e_{b}) = K_{ab} {^{i}}
\phi_{i}, \quad J_{ai} = g (D_{\delta} e_{a}, n_{i}) =
\widetilde{\nabla}_{a} \phi_{i};
\end{equation}
similarly, the deformation in the induced metric on {\it m} is
given by
\begin{equation}
     D_{\delta} \gamma_{ab} = 2 \beta_{ab} = 2 K_{ab} {^{i}}
\phi_{i}, \quad D_{\delta} \gamma^{ab} = - 2 \beta^{ab}.
\end{equation}
For the case treated here, this is sufficient about the
deformations of the intrinsic geometry. \\[2em]

\noindent{\uno 2.3 Deformations of the extrinsic geometry}
\vspace{.5cm}

Introducing a covariant deformation derivative as
$\widetilde{D}_{\delta} \Psi_{i} \equiv D_{\delta} \Psi_{i} -
\gamma_{i}{^{j}} \Psi_{i}$, where $\gamma_{ij} = g(D_{\delta}
n_{i}, n_{j}) = - \gamma_{ij}$, the covariant measure of the
deformations of the quantities characterizing the extrinsic
geometry are given by
\begin{eqnarray}
    \!\! & & \!\! D_{s} n_{i} = - J_{ai} e^{a} + \gamma_{ij}
n^{j}, \quad \widetilde{D}_{\delta} n_{i} = - J_{ai} e^{a} =
- (\widetilde{\nabla}_{a} \phi_{i}) e^{a}, \\
     \!\! & & \!\! \widetilde{D}_{\delta} K_{ab} {^{i}} =
- \widetilde{\nabla}_{a} \widetilde{\nabla}_{b} \phi^{i} + [K_{ac}
{^{i}} K^{c} {_{bj}}-g(R(e_{a}, n_{j}) e_{b}, n^{i})]
\phi^{j}, \\
     \!\! & & \!\! \widetilde{D}_{\delta} \omega_{a} {^{ij}}
- \nabla_{a} \gamma^{ij} =  D_{\delta} \omega_{a} {^{ij}} -
\widetilde{\nabla}_{a} \gamma^{ij} = - K_{ab} {^{i}}
\widetilde{\nabla}^{b} \phi^{j} + K_{ab} {^{j}}
\widetilde{\nabla}^{b} \phi^{i}  \nonumber \\
     \!\! & & \!\! \hspace{6.5cm} + g(R(n_{k}, e_{a}) n^{j}, n^{i})
\phi^{k},
\end{eqnarray}
which depend on second derivatives of $\phi_{i}$; the notation
$g(R(Y_{1}, Y_{2})Y_{3}, Y_{4}) = R_{\mu\nu\alpha\beta}
Y^{\nu}_{1} Y^{\mu}_{2} Y^{\alpha}_{3} Y^{\beta}_{4}$ is used,
where $R_{\mu\nu\alpha\beta}$ is the Riemann tensor of spacetime.
Other useful formulae and more details can be found directly in
Ref.\ \cite{6}. \\[2em]

\noindent {\uno III. THE DIRAC-NAMBU-GOTO ACTION AND ITS
DEFORMATIONS}
\vspace{1em}

The simplest phenomenological theory of a topological defect is
given by the most simple generally covariant action, proportional
the area swept out by the worldsheet:
\begin{equation}
     S = - \sigma \int d^{D} \xi \sqrt{-\gamma},
\end{equation}
where $\sigma$ is the brane tension. If we restrict us to closed
defects without physical boundaries, the corresponding equations
of motion are given by
\begin{equation}
     \Delta X^{\mu} + \Gamma^{\mu}_{\alpha\beta} (X^{\nu})
\gamma^{ab} e^{\alpha}_{a} e^{\beta}_{b} = 0,
\end{equation}
where $\Delta = \frac{1}{\sqrt{-\gamma}} \partial_{a}
(\sqrt{-\gamma} \gamma^{ab} \partial_{b})$, and
$\Gamma^{\mu}_{\alpha\beta} (X^{\nu})$ are the spacetime
Christoffel symbols evaluated on {\it m}. Because all but $N-D$
linear combinations of these equations are identically satisfied,
they are entirely equivalent to the $N-D$ equations
\begin{equation}
     K^{i} = \gamma^{ab} K_{ab} {^{i}} = 0,
\end{equation}
which describe extremal surfaces. Furthermore, using Eqs.\ (5),
(7), and (11), one can find the deformation equations of motion in
terms of the scalar fields $\phi^{i}$ \cite{6}:
\begin{eqnarray}
     - \widetilde{D}_{\delta} K^{i} \!\! & = & \!\! - \gamma^{ab}
\widetilde{D}_{\delta} K_{ab}{^{i}} - K_{ab}{^{i}} D_{\delta}
\gamma^{ab} \nonumber \\
     \!\! & = & \!\! [\widetilde{\Delta}\delta^{ij}- (M^{2})^{ij}]\phi_{j} = 0,
\end{eqnarray}
where the d'Alembertian $\widetilde{\Delta} =
\widetilde{\nabla}^{a} \widetilde{\nabla}_{a}$, and the effective
mass matrix $(M^{2})^{ij}=-K_{ab}{^{i}}K^{ab} {^{j}} - g (R(e_{a},
n^{j}) e^{a}, n^{i})$. Note that the linear operator ${\cal
E}^{ij}\equiv \widetilde{\Delta}\delta^{ij}- (M^{2})^{ij}$ takes
vector fields into themselves, and the mass matrix is symmetric,
$(M^{2})^{ij}=(M^{2})^{ji}$. By direct substitution of the
relation $\phi^{i} = n^{i}_{\mu} \delta X^{\mu}$ into Eq.\ (12),
it is a straightforward matter to write down the corresponding
equations for spacetime variations $\xi^{\mu}$:
\[
     n^{i}_{\mu} D^{a} D_{a} \xi^{\mu} + (\widetilde{\nabla}^{a}
n^{i}_{\mu}) D_{a} \xi^{\mu} + [ h^{\alpha\beta}
\bot^{\lambda}{_{\mu}} R^{\sigma} {_{\alpha\lambda\beta}}
n^{i}_{\sigma} - 2 \omega^{a}{_{ij}} \widetilde{\nabla}_{a}
n^{j}_{\mu}] \xi^{\mu} = 0,
\]
where we have exploited the projection tensor $h^{\mu\nu} \equiv
\gamma^{ab} e^{\mu}_{a} e^{\nu}_{b} = g^{\mu\nu} - \bot^{\mu\nu}$,
with $\bot^{\mu\nu} = n^{\mu}_{i} n^{i\nu}$. The normal projection
of previous equation on $n_{\nu i}$, gives finally
\begin{equation}
     ({\cal E} \xi^{\mu})_{\nu} \equiv [ \bot_{\mu\nu} D^{a} D_{a}
+ (n_{\nu i} \widetilde{\nabla}^{a} n^{i}_{\mu}) D_{a} +
h^{\alpha\beta} \bot^{\sigma}{_{\nu}} \bot^{\lambda}{_{\mu}}
R_{\sigma\alpha\lambda\beta} - 2 \omega^{a}{_{ij}} n_{\nu}{^{i}}
\widetilde{\nabla}_{a} n^{j}_{\mu} ] \xi^{\mu} = 0;
\end{equation}
similarly the linear operator $\cal E$ is taking vector fields
into themselves. As we shall see in Section 5.2, one can to find
from both equations (12) and (13), a local continuity equation on
the phase space. \\[2em]

\noindent {\uno IV. COVARIANT PHASE SPACE AND THE EXTERIOR
CALCULUS}
\vspace{1em}

In according to Witten \cite{1,3}, in a given physical theory,
{\it the classical phase space is the space of solutions of the
classical equations of motion}, which corresponds to a manifestly
covariant definition. The basic idea of the covariant description
of the canonical formalism is to construct a symplectic structure
on such a phase space, instead of choosing {\it p's} and {\it
q's}.

In the present case, the phase space is the space of solutions of
Eqs.\ (10) (or equivalently Eqs.\ (11)), and we shall call it $Z$.
Any background quantity, such as those defined in Section 2.1,
will be associated with zero-forms on $Z$. The deformation
operator $\delta$ acts as an exterior derivative on $Z$, taking
$k$-forms into $(k+1)$-forms, and it should satisfy
\begin{equation}
     \delta^{2} = 0,
\end{equation}
and the Leibniz rule
\begin{equation}
     \delta (AB) = \delta AB + (-1)^{A} A \delta B.
\end{equation}
In particular, $\delta X^{\mu}$ is the exterior derivative of the
zero-form $X^{\mu}$, and it will be closed:
\begin{equation}
     \delta^{2} X^{\mu} = 0.
\end{equation}
Furthermore, since $\phi^{i} = n^{i}_{\mu} \delta X^{\mu}$, and
$n^{i}_{\mu}$ corresponds to a zero-forms on $Z$, the scalar
fields $\phi^{i}$ are one-forms on $Z$, and thus are
anticommutating objects: $\phi^{i}\phi^{j} = - \phi^{j}\phi^{i}$.
This property allows us to verify that, being the vector field
$\delta = n^{i} \phi_{i}$, thus $\delta^{2} = n^{i} n^{j}
\phi_{i}\phi_{j}$, which vanishes because of the commutativity of
the zero-forms $n^{i}$ and the anticommutativity of the $\phi^{i}$
on $Z$, in fully agreement with Eq.\ (14). It is important to
mention, at this point, that the covariant deformation operator
$D_{\delta}$ (and subsequently $\widetilde{D}_{\delta}$) also
works as an exterior derivative on $Z$, in the sense that maps
$k$-forms into $(k+1)$-forms; however $D^{2}_{\delta}$ does not
vanish necessarily. In this manner, from Eq.\ (4) we can identify
$\beta_{ab}$ and $J_{aj}$ as one-forms on $Z$, and similarly for
$\gamma_{ij}$ in Section 2.3.

Which these preliminary, we can determine certain two-forms on $Z$
that will be useful for our present proposes. Considering that
$\delta \equiv \delta X^{\mu} \partial_{\mu}$, and $D_{\delta}
\equiv \delta X^{\mu} D_{\mu}$, we can show that $D_{\delta}
(\delta X^{\mu})$ vanishes:
\begin{equation}
     D_{\delta} (\delta X^{\mu}) = \delta X^{\alpha} D_{\alpha}
\delta X^{\mu} = \delta X^{\alpha} [ \partial_{\alpha} \delta
X^{\mu} + \Gamma^{\mu}_{\alpha\lambda} \delta X^{\lambda}] =
\delta^{2} X^{\mu} + \Gamma^{\mu}_{\alpha\lambda} \delta
X^{\alpha} \delta X^{\lambda} = 0,
\end{equation}
where the first term vanishes in according to Eq.\ (16), and the
second one because of the symmetry of
$\Gamma^{\mu}_{\alpha\lambda}$ in the indices $\alpha$ and
$\lambda$, and the anticommutativity of $\delta X^{\alpha}$ and
$\delta X^{\lambda}$. Hence, Eq.\ (17) suggests that $D_{\delta}$
is, as well as $\delta$, a measure of the closeness of $\delta
X^{\mu}$ on $Z$. Furthermore, using Eqs.\ (6), (17), and the
Leibniz rule for $D_{\delta}$, we find that:
\[
     D_{\delta} \phi^{i} = D_{\delta} (n^{i}_{\mu} \delta
X^{\mu})= D_{\delta} n^{i}_{\mu} \delta X^{\mu} + n^{i}_{\mu}
D_{\delta} \delta X^{\mu} = D_{\delta} n^{i}_{\mu} (n^{\mu}_{j}
\phi^{j}) = \gamma^{i} {_{j}} \phi^{j},
\]
which implies that
\begin{equation}
     \widetilde{D}_{\delta} \phi^{i} = 0.
\end{equation}

Although effectively $\widetilde{D}^{2}_{\delta}$ does not vanish,
Eq.\ (18) suggests that $\widetilde{D}_{\delta}$ is not only a
covariant measure of the deformations, but also the measure of the
closeness of $\phi^{i}$ on $Z$. In the CG deformation scheme, the
$\phi^{i}$ are considered, in a conventional sense, as scalar
fields ``living on the worldsheet"; in the present scheme such
fields can be considered as closed (in the sense of Eq.\ (18))
one-forms, ``living on the corresponding phase space $Z$". The
property (18) of the $\phi^{i}$ will be essential for our present
purposes. \\[2em]

\noindent {\uno V. SELF-ADJOINT OPERATORS AND A COVARIANTLY
CONSERVED CURRENT ON Z}
\vspace{1em}

In this Section we shall construct, using the concept of
self-adjoint operators, a worldsheet covariantly conserved
two-form on $Z$.
\\[2em]

\noindent {\uno 5.1. Adjoint operators and local continuity laws}
\vspace{0.5cm}

The general relationship between adjoint operators and covariantly
conserved currents has been already given in previous works (see
for example \cite{2} and references cited therein), however we
shall discuss it in this section for completeness.

If ${\cal P}$ is a linear partial differential operator which
takes matrix-valued tensor fields into themselves, then, the
adjoint operator of ${\cal P}$, is that operator ${\cal
P}^{\dag}$, such that
\begin{equation}
   {\rm Tr} \{ f^{\rho\sigma ...}[{\cal P}(g_{\mu\nu ...})]_{\rho\sigma ...}
- [{\cal P}^{\dag}(f^{\rho\sigma ...})]^{\mu\nu ...}g_{\mu\nu ...}
\} = \nabla_{\mu} {\cal J}^{\mu},
\end{equation}
where $\rm Tr$ denotes the trace and ${\cal J}^{\mu}$ is some
vector field. From this definition, if $\cal Q$ and $\cal R$ are
any two linear operators, one easily finds the following
properties:
\[
   ({\cal Q}{\cal R})^{\dag} = {\cal R}^{\dag}{\cal Q}^{\dag}, \qquad
({\cal Q} + {\cal R})^{\dag} = {\cal Q}^{\dag} + {\cal R}^{\dag},
\]
and in the case of a function $F$,
\[
    F^{\dag} = F,
\]
which will be used implicitly below.

From Eq.\ (19) we can see that this definition automatically
guarantees that, if ${\cal P}$ is a self-adjoint operator (${\cal
P}^{\dag} = {\cal P}$), and the fields $f$  and $g$ correspond to
a pair of solutions admitted by the linear system ${\cal P} (f) =
0 = {\cal P}(g)$, then we obtain the continuity law $\nabla_{\mu}
{\cal J}^{\mu}=0$, which establishes that ${\cal J}^{\mu}$ is a
covariantly conserved current, bilinear on the fields $f$ and $g$.
This fact means that for any self-adjoint homogeneous equation
system, one can always to construct a conserved current. Although
this result has been established assuming only tensor fields and
the presence of a single equation, such a result can be extended
in a direct way to equations involving spinor fields, matrix
fields, and the presence of more that one field.

In the present work, the indices appearing in Eq.\ (19), can
correspond to spacetime, internal, and/or normal indices.
Furthermore, since the fields $f$ and $g$ will be identified with
deformations (which correspond to one-forms on $Z$), the products
of the form $f {\cal P} (g)$ and ${\cal P}^{\dag} (f)g$ in Eq.\
(19) must be understood as exterior products on $Z$, and something
similar for the field ${\cal J}^{\mu}$ in its dependence on the
fields $f$ and $g$. These subjects will be clarified in the
example described below. \\[2em]

\noindent {\uno 5.2. Self-adjointness of the operators governing
the deformations} \vspace{0.5cm}

In this section we shall show that the operators ${\cal E}^{ij}$
and ${\cal E}$ in Eqs.\ (12) and (13) are indeed self-adjoint.

The case of ${\cal E}^{ij}$ is relatively simple: let
$\phi_{1}^{i}$ and $\phi_{2}^{i}$ be two scalar fields (which will
be identified as a pair of solutions of Eqs.\ (12), we mean a pair
of one-forms on $Z$), then it is very easy to prove the following
identity,
\[
   \phi_{1i}\widetilde{\Delta}\phi_{2}^{i} \equiv \phi_{1i}
\widetilde{\nabla}^{a} \widetilde{\nabla}_{a}
\phi_{2}^{i}\equiv(\widetilde{\Delta}\phi_{1i})\phi_{2}^{i} +
\nabla_{a}j^{a}, \qquad j^{a}=\phi_{1i}\widetilde{\nabla}^{a}
\phi_{2}^{i} -(\widetilde{\nabla}^{a} \phi_{1i})\phi_{2}^{i},
\]
which implies that
\begin{equation}
\phi_{1i}[\widetilde{\Delta}\delta^{ij}- (M^{2})^{ij}]\phi_{2j}=
[[\widetilde{\Delta}\delta^{ji}- (M^{2})^{ji}]\phi_{1i}]\phi_{2j}+
\nabla_{a}j^{a},
\end{equation}
where we have considered the symmetry of the mass matrix; such a
mass term does no contribute explicitly to $j^{a}$ (although it
does implicitly). Eq.\ (20) shows that, in according to Eq.\ (19),
${\cal E}^{ij}$ is self-adjoint. In this manner,  if $\phi_{1i}$
and $\phi_{2i}$ corresponds to a pair of solutions of Eqs. (12),
$\nabla_{a}j^{a}=0$, which implies that, at each brane worldsheet
point, $j^{a}$ is a covariantly conserved two-form on $Z$. This
bilinear product is evidently antisymmetric in $\phi_{1i}$ and
$\phi_{2i}$; thus, we can set $\phi_{1i}=\phi_{2i}=\phi_{i}$ ,
without loosing generality, and write simply
\begin{equation}
j^{a}= \phi_{i}\widetilde{\nabla}^{a} \phi^{i}.
\end{equation}
On the other hand, for demonstrating the self-adjointness of the
operator $\xi$ governing the deformations in Eq.\ (13), we need
the following identities which are very easy of verifying. Let
$\xi^{\nu}_{1}$ and $\xi^{\mu}_{2}$ be two spacetime vector
fields, then:
\begin{eqnarray}
     \xi^{\nu}_{1} \bot_{\mu\nu} D^{a} D_{a} \xi^{\mu}_{2} \!\! & \equiv &
\!\! \nabla_{a} [ \bot_{\mu\nu} \xi^{\nu}_{1} D^{a} \xi^{\mu}_{2}
- D^{a} (\bot_{\mu\nu} \xi^{\nu}_{1}) \xi^{\mu}_{2}] + [D^{a}
D_{a} (\bot_{\mu\nu} \xi^{\nu}_{1})] \xi^{\mu}_{2}, \nonumber \\
     \xi^{\nu}_{1} n_{\nu i} \widetilde{\nabla}^{a} n^{i}_{\mu}
D_{a} \xi^{\mu}_{2} \!\! & \equiv & \!\! \nabla_{a} [(n_{\nu i}
\widetilde{\nabla}^{a} n^{i}_{\mu}) \xi^{\nu}_{1} \xi^{\mu}_{2}] -
D_{a} [( n_{\nu i} \widetilde{\nabla}^{a} n^{i}_{\mu})
\xi^{\nu}_{1}] \xi^{\mu}_{2};
\end{eqnarray}
furthermore, using the definition $\bot_{\mu\nu} \equiv n_{\mu i}
n_{\nu} {^{i}}$, it is straightforward to show the following
background identities:
\begin{eqnarray}
     D^{a} \bot_{\mu\nu} \!\! & = & \!\! n_{\nu i}
\widetilde{\nabla}^{a} n^{i}_{\mu} + n_{\mu} {^{i}}
\widetilde{\nabla}^{a} n_{\nu i}, \nonumber \\
     D^{a} D_{a} \bot_{\mu\nu} - 2 D_{a} (n_{\nu i} \widetilde{\nabla}^{a}
n^{i}_{\mu}) \!\! & = & \!\! 2 \omega^{aij} (n_{\nu i}
\widetilde{\nabla}_{a} n_{\mu j} - n_{\mu i}
\widetilde{\nabla}_{a} n_{\nu j}).
\end{eqnarray}

In this manner, using Eqs.\ (22) and (23), and the explicit form
of the operator $\cal E$ in Eq.\ (13), one finds that
\begin{eqnarray}
     \xi^{\nu}_{1} ({\cal E} \xi^{\mu}_{2})_{\nu} \!\! & = & \!\!
\{ \bot_{\mu\nu} D^{a} D_{a} \xi^{\nu}_{1} + 2(n_{\mu i}
\widetilde{\nabla}^{a} n_{\nu} {^{i}}) \nabla_{a} \xi^{\nu}_{1} +
[-2 \omega^{aij} n_{\mu i} \widetilde{\nabla}_{a} n_{\nu j} +
h^{\alpha\beta} \bot^{\sigma} {_{\nu}} \bot^{\lambda} {_{\mu}}
R_{\sigma\alpha\lambda\beta}] \xi^{\nu}_{1} \} \xi^{\mu}_{2}
\nonumber \\
     \!\! & & \!\! + \nabla_{a} j^{a} = ({\cal E} \xi^{\nu}_{1})_{\mu}
\xi^{\mu}_{2} + \nabla_{a} j^{a},
\end{eqnarray}
which corresponds to Eq.\ (19) with ${\cal E}^{\dag} = {\cal E}$,
and
\begin{equation}
     j^{a} \equiv \bot_{\mu\nu} \xi^{\nu}_{1} \nabla^{a}
\xi^{\mu}_{2} + (n_{\nu i} \widetilde{\nabla}^{a} n^{i}_{\mu})
\xi^{\nu}_{1} \xi^{\mu}_{2} - [\bot_{\mu\nu} (\nabla^{a}
\xi^{\nu}_{1}) \xi^{\mu}_{2} + (n_{\mu} {^{i}}
\widetilde{\nabla}^{a} n_{\nu i}) \xi^{\nu}_{1} \xi^{\mu}_{2}].
\end{equation}

Similarly, if $\xi^{\nu}_{1}$ and $\xi^{\mu}_{2}$ corresponds to a
pair of solutions admitted by Eqs.\ (13), $({\cal E}
\xi^{\nu}_{1})_{\mu} = 0 = ({\cal E} \xi^{\mu}_{2})_{\nu}$, then
from Eq.\ (24), we obtain the local continuity equation:
$\nabla_{a} j^{a} = 0$. $j^{a}$  in Eq.\ (25) is  antisymmetric
in $\xi^{\nu}_{1}$ and $\xi^{\mu}_{2}$; thus, setting
$\xi^{\nu}_{1} = \xi^{\nu}_{2} = \xi^{\nu}$, $j^{a} =
\bot_{\mu\nu} \xi^{\nu} \nabla^{a} \xi^{\mu} + (n_{\nu i}
\widetilde{\nabla}^{a} n^{i}_{\mu}) \xi^{\nu}\xi^{\mu}$, which has
a remarkable simplification in terms of the fields $\phi^{i} =
n^{i}_{\mu} \xi^{\mu}$:
\begin{eqnarray}
     j^{a} \!\! & = & \!\! n_{\nu i} \xi^{\nu} [n^{i}_{\mu}
\nabla^{a} \xi^{\mu} + (\nabla^{a} n^{i}_{\mu} - \omega^{aij}
n_{\mu j}) \xi^{\mu}] \nonumber \\
     \!\! & = & \!\! \phi_{i} [n^{i}_{\mu}
\nabla^{a} \xi^{\mu} + (\nabla^{a} n^{i}_{\mu}) \xi^{\mu} -
\omega^{aij} \phi_{j}] \nonumber \\
     \!\! & = & \!\! \phi_{i} [\nabla^{a} (n^{i}_{\mu} \xi^{\mu})
- \omega^{aij} \phi_{j}] = \phi_{i} \widetilde{\nabla}^{a}
\phi^{i},\nonumber
\end{eqnarray}
in fully agreement with Eq.\ (21). Note that $j^{a}$, with support
confined to the brane worldsheet, transforms as a scalar under
normal frame rotations. In next Section we shall discuss about the
closeness of this bilinear form on $Z$. \\[2em]

\noindent {\uno VI. THE SYMPLECTIC STRUCTURE ON Z}
\vspace{1em}

Strictly speaking, a (non-degenerate) closed two-form on $Z$ is
called a symplectic structure; the closeness holds if the exterior
derivative of such a two-form vanishes. This property is
equivalent to that Poisson brackets satisfy, in the usual
Hamiltonian scheme, the Jacoby identity.

Hence, let us determine the exterior derivative of the two-form
$j^{a}$ previously constructed in Eq.\ (21). We calculate first
$\widetilde{D}_{\delta} \widetilde{\nabla}_{b} \phi^{i}$; for
which we can follow the calculation for obtaining Eq.\ (4.15) in
Ref.\ \cite{6}, and to consider, of course, that in the present
approach $\widetilde{D}_{\delta}$ acts as an exterior derivative
on $Z$, and, in particular, obeys the Leibniz rule (15):
\begin{eqnarray}
     \widetilde{D}_{\delta} \widetilde{\nabla}_{b} \phi^{i} \!\! & = &
\!\! D_{\delta} [D_{b} \phi^{i} - \omega_{b} {^{ij}} \phi_{j}] -
\gamma^{ij} \widetilde{\nabla}_{b} \phi_{j} \nonumber \\
     \!\! & = & \!\! D_{b} D_{\delta} \phi^{i} - (D_{\delta} \omega_{b}
{^{ij}}) - \omega_{b} {^{ij}} D_{\delta} \phi_{j} -
\widetilde{\nabla}_{b} (\gamma^{ij} \phi_{j}) +
(\widetilde{\nabla}_{b} \gamma^{ij}) \phi_{j} \nonumber \\
     \!\! & = & \!\! \widetilde{\nabla}_{b} \widetilde{D}_{\delta}
\phi^{i} - (D_{\delta} \omega_{b} {^{ij}} - \widetilde{\nabla}_{b}
\gamma^{ij}) \phi_{j}, \nonumber
\end{eqnarray}
the first term vanishes in according to Eq.\ (18), whereas the
second term can be rewritten using Eq.\ (8),
\begin{eqnarray}
     \widetilde{D}_{\delta} \widetilde{\nabla}_{b} \phi^{i} \!\! & = &
\!\! [K_{bc} {^{i}} J^{cj} - K_{bc} {^{j}} J^{ci} - g (R(n_{k},
e_{b}) n^{j}, n^{i}) \phi^{k}] \phi_{j} \nonumber \\
     \!\! & = & \!\! - K_{bc} {^{i}} j^{c} + \beta_{bc} J^{ci} - g (R(n_{k},
e_{b}) n^{j}, n^{i}) \phi^{k} \phi_{j},
\end{eqnarray}
in the last line, we have considered that, in according to Eq.\
(21), $j^{a} = \phi_{i} J^{ai} = - J^{ai} \phi_{i}$ ($J^{ai}$ and
$\phi_{i}$ are one-forms), and $K_{bc} {^{j}} \phi_{j} =
\beta_{bc}$ (see first of Eq.\ (4)). Now, it is very easy to find
that:
\[
     D_{\delta} j_{b} = \widetilde{D}_{\delta} j_{b} =
\widetilde{D}_{\delta} (\phi^{i} \widetilde{\nabla}_{b} \phi_{i})
= (\widetilde{D}_{\delta} \phi^{i}) \widetilde{\nabla}_{b}
\phi_{i} - \phi^{i} \widetilde{D}_{\delta} \widetilde{\nabla}_{b}
\phi_{i},
\]
where we have used again the Leibniz rule, and one more time the
first term vanishes according to Eq.\ (18), and using Eq.\ (26)
the second term reduces to
\begin{equation}
     D_{\delta} j_{b} = 2 \beta_{bc} j^{c} + g (R(n_{k}, e_{b}) n_{j},
n_{i}) \phi^{i} \phi^{k} \phi^{j}.
\end{equation}
Thus, considering Eq.\ (27), and the second of Eq.\ (5), one finds
that
\begin{eqnarray}
     \delta j^{a} = D_{\delta} j^{a} = D_{\delta} (\gamma^{ab}
j_{b}) \!\! & = & \!\! (D_{\delta} \gamma^{ab}) j_{b} +
\gamma^{ab} D_{\delta} j_{b} \nonumber \\
     \!\! & = & \!\! e^{a\nu} R_{\mu\nu\alpha\beta} n^{\mu}_{k}
n^{\beta}_{j} n^{\alpha}_{i} \phi^{i} \phi^{k} \phi^{j}
\nonumber \\
     \!\! & = & \!\! e^{a\nu} R_{\mu\nu\alpha\beta} \xi ^{\alpha}
\xi^{\mu} \xi^{\beta}.
\end{eqnarray}
However, the projection of the spacetime Riemann tensor on the
right-hand side of Eq.\ (28) vanishes, since considering the
definitions $2 D_{[\mu} D_{\nu ]} A_{\alpha} = R_{\mu\nu\alpha}
{^{\rho}} A_{\rho}$ and $D_{\delta} \equiv \xi^{\alpha}
D_{\alpha}$, it can be rewritten as
\begin{equation}
     R_{\mu\nu\alpha\beta} \xi^{\alpha} \xi^{\mu} \xi^{\beta} = -
2 \xi^{\alpha} D_{\delta} D_{\alpha} \xi_{\nu} = 2 D_{\delta}
(D_{\delta} \xi_{\nu}) = 0,
\end{equation}
where Eqs.\ (15) and (17) has been considered. Therefore, the
symplectic current is closed on $Z$,
\begin{equation}
     \delta j^{a} = 0.
\end{equation}

Since we have explicitly assumed that the worldsheet is
orientable, we can define integration in a direct way, and to
construct the following two-form in terms of $j^{a}$, which will
correspond finally to the wanted symplectic structure:
\begin{equation}
     \omega \equiv \int_{\Sigma} \sqrt{-\gamma}j^{a} d \Sigma_{a},
\end{equation}
where ${\Sigma}$ is a spacelike section of $m$, and corresponds to
an initial value (hyper)surface for the configuration of the
defect. Since $j^{a}$ is covariantly conserved, $\omega$ in Eq.\
(31) is independent of the choice of ${\Sigma}$. On the other
hand, since $\delta\sqrt{-\gamma}=0$ (from this condition the
equations of motion for the defect are obtained from Eq.\ (9)),
the closeness of $\omega$ holds if  $j^{a}$ itself is closed, and
then, from Eqs.\ (30), and (31), $\delta\omega = 0$. Hence,
$\omega$ is the wanted symplectic structure on $Z$.

It remains to discuss gauge invariance of $\omega$. Since all
fields appearing in the definition of $\omega$ transform
homogeneously, like tensors, $\omega$ is invariant under spacetime
diffeomorphisms. Similarly, since $\omega$ involves integration of
a worldsheet density, $\omega$ is also invariant under worldsheet
reparametrizations. With respect to normal frame rotations,
$\phi_{i}$ and $\widetilde{\nabla}^{a} \phi^{i}$ transform
homogeneously, like vectors, and then $j^{a}$ and $\omega$,
transforming as scalars, are invariant under such
rotations. \\[2em]

\noindent {\uno VII. DISCUSSIONS AND OPEN QUESTIONS}
\vspace{1em}

In this manner, we have obtained a symplectic structure on $Z$
for defects propagating on a curved background spacetime, and it
emerges in a natural way and without additional assumptions.
However, as pointed out in Refs.\ \cite{6,7,8}, the CG scheme
employed in the present treatment fails to treat the topological
defect as a source for gravity, which should be considered in a
more complete description of deformations, and in the
construction of the corresponding symplectic structure. Within
this complete scheme, the deformations of the ambient spacetime
itself will be taken into account, and it is possible that
similar results can be obtained in such situation. We hope to
address this subject in forthcoming works.

As a by-product, the worldsheet conserved current $j^{a}$ obtained
in Section 5.2, and used for generating a symplectic structure on
$Z$ in the present approach, can be considered, in a more ordinary
sense, as a Noetherian current that allows us to obtain conserved
currents associated with any continuous symmetries of the
background.

It is opportune to comment on the role that the deformation gauge
connection $\gamma^{ij}$ plays on the phase space $Z$. Such as in
the CG scheme, in the present approach it never appears explicitly
in any relevant physical quantity on $Z$, such as $j^{a}$ and
$\omega$, but it shows up in intermediate calculations, for
example in Eqs.\ (18) and the first one of the previous section.
Therefore, the connection neither needs to be calculated
explicitly on $Z$.

Finally, in spite of we have considered the simplest theory for an
extended topological defect, the present results show that the
symplectic geometry of the phase space possesses a rich underlying
structure, deserving a more wide investigation. In fact, using a
different approach for the geometry of deformations of defects
given by Carter, we have obtained results analogous to those
presented here \cite{9}, and questions such as the existence of a
symplectic potential, degenerate directions of the
(pre-)symplectic structure, are explicitly discussed \cite{10}.
These results will provide a very powerful geometrical approach
for the quantization of such objects, and we hope to extend our
achievements elsewhere. \\[2em]

\begin{center}
{\uno ACKNOWLEDGMENTS}
\end{center}
\vspace{1em}

This work was supported by CONACyT and the Sistema Nacional de
Investigadores (M\'exico). The author wants to thank H.
Garcia Compean for drawing my attention on this matter. \\[2em]

}

\begin{thebibliography}{}
\setlength{\itemsep}{-.50em}
\bibitem{1} C. Crncovi\'c and E. Witten, in {\it Three Hundred Years
of Gravitation}, edited by S. W. Hawking and W. Israel (Cambridge
University Press. Cambridge, 1987); A. Ashtekar, L. Bombelli, O.
Reula in {\it Mechanics, Analysis and Geometry: 200 Years after
Lagrange}, Francaviglia Ed., Elsevier Science Publisher (1991);
Lee  J., R. M. Wald, J.\ Math.\ Phys.\ {\bf 31},725 (1990).
\bibitem{2} R. Cartas-Fuentevilla  J.\ Math.\ Phys.\ {\bf 43}, 644
(2002).
\bibitem{3} E. Witten, Nucl.\ Phys.\ {\bf B276}, 291 (1986).
\bibitem{4} M. Chu, P. Goddard, I. Halliday, D. Olive, and
Schwimmer, Phys.\ Lett.\ B {\bf 266}, 71 (1991).
\bibitem{5} K. S. Soh, Phys.\ Rev.\ D {\bf 49}, 1906 (1994).
\bibitem{6} R. Capovilla and J. Guven, Phys.\ Rev.\ D {\bf 51},
6736 (1995).
\bibitem{7} R. Capovilla and J. Guven, Phys.\ Rev.\ D {\bf 57},
5158 (1998).
\bibitem{8} J. Guven, Phys.\ Rev.\ D {\bf 48}, 5562 (1993).
\bibitem{9} R. Cartas-Fuentevilla, {\it ''Identically closed two-form for phase
space quantization of Dirac-Nambu-Goto p-branes in a curved
spacetime"}, to be published, Phys. Lett. B (2002).
\bibitem{10} R. Cartas-Fuentevilla, {\it ''Global symplectic
potentials on the Witten covariant
phase space for bosonic extendons"}, submitted to Phys. Lett. B. (2002).
\end{thebibliography}
\end{document}